\documentstyle[rotate, psfig]{mn}

\title[Time delay of B1422+231]
{ 
Determination of time delay from the gravitational lens B1422+231
      }

\author[A.R.~Patnaik \& D.~Narasimha]
{A.R.~Patnaik$^1$, D.~Narasimha$^2$\\
$^1$Max-Planck-Institut f{\"u}r Radioastronomie, 
Auf dem H{\"u}gel 69, D--53121 Bonn, FRG \\
email: apatnaik@mpifr-bonn.mpg.de\\
$^2$Department of Astronomy and Astrophysics, TIFR, Homi Bhabha
Road, Mumbai 400005, India.\\
email: dna@astro.tifr.res.in\\
}

\begin{document}

\maketitle

\begin{abstract}
  
  We present the radio light curves of lensed images of the
  gravitational lens B1422+231. The observations have been carried out
  using the VLA at 8.4 and 15~GHz over a period of 197 days. We
  describe a method to estimate the time delay from the observed light
  curves.  Using this method, our cross-correlation analysis shows
  that the time delay between images B and A is 1.5$\pm$1.4d,
  between A and C is 7.6$\pm$2.5d, between B and C is
  8.2$\pm$2.0d. When applied to  other lensed systems with measured
  time delays our new method gives comparable
  results.

\end{abstract}


\section{Introduction}

One of the most promising outcomes of the study of gravitational
lensing is the potential to determine the Hubble constant, H$_0$,
independent of the standard distance ladder (Refsdal 1964). The light
travel times for the lensed images are not equal due to the geometrical
path length difference caused by the gravitational potential of the
lens. This results in a time delay between the lensed images that can
be measured by monitoring the images over a period of time.  Thus,
given a measurement of time delay and a knowledge of the gravitational
potential of the lens, one can estimate H$_0$. At present, the
accuracy of H$_0$ values determined by this method is limited mainly
by our knowledge of the mass
distribution in the distant lensing galaxy.

So far, time-delay measurements are available for seven lensed systems.
For the twin quasar 0957+561 the measured values at both radio and
optical wavelengths are in agreement and have been confirmed by observing
variability in the image B after it was recorded in the image A
(Schild \& Thomson 1997, Kundi\'c et al. 1997b, Haarsma et al. 1999).
While for this system a reliable measurement of the time delay has
been achieved, translation into an estimate of the Hubble constant
turns out to be not straightforward (Narasimha, 1999).  For
the systems PG~1115+080 (Schechter et al. 1997), B1608+656 (Fassnacht
et al. 1999), PKS~1830-211 (Lovell et al. 1998), B1600+434 (Koopmans
et al. 2000) the measured delays are less accurate, mainly because
it is not easy to extract the intrinsic variability and determine its
phase lag between the images.  The system B0218+357 has good time delay
measurements (Biggs et al. 1999) and the redshift combination of the
lens and source are favourable for a possible determination of
H$_0$, but the lens models continue to be
ill-constrained.  However, these systems together cover various
combinations of the distance to the source and lens as well as the
distance from the lens to the source; hence an estimate of the
distance scale from time-delays in all these systems together would be
an important step to determine the Hubble constant and the large scale
geometry of the Universe.

The lens system B1422+231 is particularly interesting in this respect,
as the redshift of the source, z$_S = 3.62$, (Patnaik et al. 1992) is
very large compared to the redshift of the lens, z$_L = 0.33$,
(Kundi\'c et al. 1997a, Tonry 1998) and consequently, the time-delay
is a more direct measure of the distance to the lens.  B1422+231 is a
quadruple system which has been observed at various wavebands
(Patnaik et al. 1992; Lawrence et al. 1992; Remy et al. 1993; Yee \&
Ellingson 1994; Bechtold \& Yee 1995; Akujor et al. 1996; Yee \&
Bechtold 1996; Impey et al. 1996). Several models have been presented
to explain the lens system (Hogg \& Blandford 1994, Narasimha \&
Patnaik, 1994, Kormann, Schneider \& Bartlemann 1994, Mao \& Schneider
1998).  Detailed radio maps at sub-milliarcsecond resolution are
available for the images (Patnaik et al. 1999) and the mapping of the
resolved structures of the images can be used for construction of the lens
models. Optical spectra of the images have been taken with high
sensitivity (Impey et al. 1996) and the emission line profiles of
strong lines like Ly$\alpha$ have been used by Narasimha \& Srianand
(1999) to constrain the lens models. Thus, it is clear that a
measurement of the time-delay for this system will be useful for the
calibration of the cosmic distance scale, even though the source is
not highly variable.

\section{Observations and Data Analysis}

The VLA monitoring campaign of the gravitational lens B1422+231 was
carried out at two frequencies, 8.4 and 15~GHz between 1994 March 03
to September 16. We used two 50~MHz-wide IF bands at each frequency.

We used 3C286 (J1331+3030) to calibrate the flux density scale, OQ208
(J1407+2827) as the phase calibrator, and 3C287 (J1330+2509) as a
control source since it is not expected to be a variable source.
The other reason for selecting this source is that it is close to the
declination of our target source to ensure proper calibration of gain
as a function of elevation.

The observations were carried out for about 30~min on each day, the
integration time for the target source was 5~min at each frequency and
the rest of the time was used for observing the calibration and
control sources.  Accurate calibration is of vital importance in this
kind of monitoring observations. The observations of the two
calibrators ensure that any variations in them must be due to calibration
errors and hence these variations can, at least partially, 
be removed from the light curve of the lensed images.

We chose two different frequencies in order to have confidence in the
measurements of time delay as it is independent of frequency, though
the variability of quasars does depend on the frequency in the sense
that they tend to be more variable at higher frequencies.  In the
present case, we chose 8.4 and 15~GHz for the following reasons.  In
the A configuration of the VLA the resolutions are about 0.2 and
0.13~arcsec at 8.4 and 15~GHz respectively, which allows clean
separation of the images.  In the B configuration, however, the two
brighter images, A and B, are blended together at 8.4~GHz since the
resolution is three times poorer, while there is no such difficulty at
15~GHz. We can sample the light curve over the periods of A and B
configurations of the VLA at 15~GHz but only during the A
configuration at 8.4~GHz.

The data were calibrated and imaged using the {\sc AIPS} software
package of National Radio Astronomy Observatory. Assumed flux
densities of 3C286 were 5.145~Jy and 5.124~Jy at the centre
frequencies of the two IF bands which are 8.4149~GHz and 8.4649~GHz
respectively, and 3.449~Jy and 3.457~Jy at 14.9649~GHz and 14.9149~GHz
respectively. Since 3C286 is partially resolved at these frequencies,
we followed the standard procedure in using the shortest baselines to
set the scale of flux density.

We assumed that both OQ208 and 3C287 are unresolved sources and used
them to determine the gain solutions. We then used the {\sc AIPS}
program GETJY to determine the flux densities of these two sources
which are bootstrapped from 3C286.  The source 3C287 is, in fact,
partially resolved in the A-array (its flux density changes from
1.4~Jy at short spacings to 0.9~Jy at the longest spacing at 15GHz).
We could have mapped the source to determine its flux density.
However, the assumption that it is a point source affects its flux
density measurement by only 0.5\% (our method gets less flux density).
Thus the assumption that 3C287 behaves as a point source is justified.
Moreover, we avoid introducing errors as a result of the mapping process.

The VLA configuration was changed from A to B during our monitoring
campaign. In fact, the data presented for 8.4~GHz was only for the
duration of the A-array. We do not see any effects of the change of
the array configuration in the light curves of the calibrators at
15GHz, giving confidence in the calibration method followed.  We used
the gain solutions for the phase calibrator, OQ208, to interpolate the
amplitude and phase corrections for the target source.

The observations were scheduled in intervals of between 2 and 11 days.
The measured flux densities are presented in Tables 1 and 2 for 8.4
and 15~GHz respectively.  At 15~GHz we had 51 epochs covering both A
and B configurations of the VLA. At 8.4~GHz we had only 18 epochs
since its span was limited to the A configuration.

\begin{table*}
\begin{center}
\begin{tabular}{lllllllllllll}

 Day & 3C287 & error  & OQ208 & error & A & error      & B & error &
 C & error  & D    & error \\
     &       &        &       &       &   &            &   &       &
   &        &      &        \\ 
    0 & 2.1288 &  $\pm$0.0037 & 1.8110 &  $\pm$0.0041 & 0.1464 & $\pm$0.0003 & 0.1565 & $\pm$0.0003 & 0.0803 & $\pm$0.0003 & 0.0041 & $\pm$0.0003\\
   1 & 2.1490 &  $\pm$0.0041 & 1.8104 &  $\pm$0.0037 & 0.1467 & $\pm$0.0002 & 0.1569 & $\pm$0.0002 & 0.0802 & $\pm$0.0002 & 0.0040 & $\pm$0.0002\\
  11 & 2.1233 &  $\pm$0.0031 & 1.7412 &  $\pm$0.0516 & 0.1457 & $\pm$0.0005 & 0.1559 & $\pm$0.0005 & 0.0804 & $\pm$0.0005 & 0.0045 & $\pm$0.0005\\
  15 & 2.1407 &  $\pm$0.0080 & 1.8119 &  $\pm$0.0072 & 0.1462 & $\pm$0.0002 & 0.1557 & $\pm$0.0002 & 0.0804 & $\pm$0.0002 & 0.0043 & $\pm$0.0002\\
  18 & 2.1347 &  $\pm$0.0087 & 1.8069 &  $\pm$0.0067 & 0.1467 & $\pm$0.0002 & 0.1571 & $\pm$0.0002 & 0.0795 & $\pm$0.0002 & 0.0045 & $\pm$0.0002\\
  21 & 2.1450 &  $\pm$0.0168 & 1.8175 &  $\pm$0.0138 & 0.1484 & $\pm$0.0002 & 0.1584 & $\pm$0.0002 & 0.0817 & $\pm$0.0002 & 0.0044 & $\pm$0.0002\\
  25 & 2.1616 &  $\pm$0.0184 & 1.8463 &  $\pm$0.0165 & 0.1505 & $\pm$0.0004 & 0.1620 & $\pm$0.0004 & 0.0813 & $\pm$0.0004 & 0.0036 & $\pm$0.0004\\
  28 & 2.1743 &  $\pm$0.0204 & 1.8335 &  $\pm$0.0129 & 0.1492 & $\pm$0.0002 & 0.1601 & $\pm$0.0002 & 0.0819 & $\pm$0.0002 & 0.0045 & $\pm$0.0002\\
  31 & 2.1392 &  $\pm$0.0079 & 1.8070 &  $\pm$0.0071 & 0.1460 & $\pm$0.0002 & 0.1574 & $\pm$0.0002 & 0.0799 & $\pm$0.0002 & 0.0040 & $\pm$0.0002\\
  36 & 2.1355 &  $\pm$0.0090 & 1.8066 &  $\pm$0.0058 & 0.1504 & $\pm$0.0006 & 0.1595 & $\pm$0.0006 & 0.0811 & $\pm$0.0006 & 0.0047 & $\pm$0.0006\\
  38 & 2.1351 &  $\pm$0.0073 & 1.8020 &  $\pm$0.0067 & 0.1475 & $\pm$0.0010 & 0.1569 & $\pm$0.0010 & 0.0803 & $\pm$0.0010 & 0.0042 & $\pm$0.0014\\
  40 & 2.1273 &  $\pm$0.0050 & 1.8211 &  $\pm$0.0041 & 0.1490 & $\pm$0.0003 & 0.1584 & $\pm$0.0003 & 0.0814 & $\pm$0.0003 & 0.0043 & $\pm$0.0003\\
  46 & 2.1352 &  $\pm$0.0054 & 1.8197 &  $\pm$0.0047 & 0.1457 & $\pm$0.0004 & 0.1556 & $\pm$0.0004 & 0.0797 & $\pm$0.0004 & 0.0042 & $\pm$0.0005\\
  51 & 2.1724 &  $\pm$0.0047 & 1.8260 &  $\pm$0.0063 & 0.1518 & $\pm$0.0004 & 0.1596 & $\pm$0.0004 & 0.0822 & $\pm$0.0004 & 0.0042 & $\pm$0.0003\\
  53 & 2.0575 &  $\pm$0.0144 & 1.8130 &  $\pm$0.0104 & 0.1480 & $\pm$0.0004 & 0.1579 & $\pm$0.0004 & 0.0800 & $\pm$0.0004 & 0.0039 & $\pm$0.0004\\
  58 & 2.1527 &  $\pm$0.0043 & 1.8227 &  $\pm$0.0067 & 0.1489 & $\pm$0.0012 & 0.1581 & $\pm$0.0012 & 0.0821 & $\pm$0.0012 & 0.0069 & $\pm$0.0011\\
  63 & 1.9588 &  $\pm$0.1552 & 1.6580 &  $\pm$0.1307 & 0.1338 & $\pm$0.0002 & 0.1416 & $\pm$0.0002 & 0.0724 & $\pm$0.0002 & 0.0040 & $\pm$0.0002\\
  68 & 1.9059 &  $\pm$0.1101 & 1.6057 &  $\pm$0.0928 & 0.1249 & $\pm$0.0014 & 0.1313 & $\pm$0.0014 & 0.0652 & $\pm$0.0013 & 0.0038 & $\pm$0.0015\\

\end{tabular}
\end{center}
\caption{ Flux densities (in Jy) of the observed sources and their
   errors at 8.4~GHz. The first column refers to the days starting
  from 1994 March 3. 
  }
\end{table*}

\begin{table*}
\begin{center}
\begin{tabular}{llllllllllllll}

 Day & 3C287 & error  & OQ208 & error & A & error      & B & error &
 C & error  & D    & error & Elevation\\
     &       &        &       &       &   &            &   &       &
   &        &      &      &  \\ 
    0 & 1.4100 &  $\pm$0.0064 &  .9243 &  $\pm$0.0055 & 0.0893 &
 $\pm$0.0008 & 0.0962 & $\pm$0.0008 & 0.0491 & $\pm$0.0008 & 0.0024 &
 $\pm$0.0008 & 76\\
   1 & 1.3844 &  $\pm$0.0048 &  .9213 &  $\pm$0.0048 & 0.0886 &
 $\pm$0.0005 & 0.0954 & $\pm$0.0005 & 0.0484 & $\pm$0.0005 & 0.0030 &
 $\pm$0.0005 & 54\\
  11 & 1.3879 &  $\pm$0.0042 &  .9130 &  $\pm$0.0039 & 0.0865 &
 $\pm$0.0005 & 0.0931 & $\pm$0.0005 & 0.0476 & $\pm$0.0005 & 0.0029 &
 $\pm$0.0006 & 63\\
  15 & 1.3701 &  $\pm$0.0215 &  .9089 &  $\pm$0.0105 & 0.0854 &
 $\pm$0.0006 & 0.0918 & $\pm$0.0006 & 0.0474 & $\pm$0.0006 & 0.0035 &
 $\pm$0.0008 & 78 \\
  18 & 1.3742 &  $\pm$0.0101 &  .8953 &  $\pm$0.0068 & 0.0855 &
 $\pm$0.0009 & 0.0927 & $\pm$0.0009 & 0.0456 & $\pm$0.0009 & 0.0030 &
 $\pm$0.0012 & 78 \\
  21 & 1.3752 &  $\pm$0.0153 &  .9074 &  $\pm$0.0072 & 0.0861 &
 $\pm$0.0006 & 0.0917 & $\pm$0.0006 & 0.0473 & $\pm$0.0006 & 0.0025 &
 $\pm$0.0006 & 74\\
  25 & 1.3501 &  $\pm$0.0142 &  .9134 &  $\pm$0.0038 & 0.0855 &
 $\pm$0.0007 & 0.0938 & $\pm$0.0007 & 0.0470 & $\pm$0.0007 & 0.0024 &
 $\pm$0.0005 & 44\\
  28 & 1.4326 &  $\pm$0.0419 &  .9462 &  $\pm$0.0174 & 0.0910 &
 $\pm$0.0007 & 0.0979 & $\pm$0.0007 & 0.0487 & $\pm$0.0007 & 0.0023 &
 $\pm$0.0010 & 60\\
  31 & 1.3855 &  $\pm$0.0142 &  .9069 &  $\pm$0.0102 & 0.0868 &
 $\pm$0.0004 & 0.0917 & $\pm$0.0004 & 0.0481 & $\pm$0.0004 & 0.0022 &
 $\pm$0.0004 & 79\\
  36 & 1.3912 &  $\pm$0.0144 &  .9009 &  $\pm$0.0082 & 0.0877 &
 $\pm$0.0007 & 0.0937 & $\pm$0.0007 & 0.0468 & $\pm$0.0007 & 0.0021 &
 $\pm$0.0006 & 78\\
  38 & 1.3784 &  $\pm$0.0123 &  .9000 &  $\pm$0.0106 & 0.0866 &
 $\pm$0.0006 & 0.0934 & $\pm$0.0006 & 0.0477 & $\pm$0.0006 & 0.0029 &
 $\pm$0.0007 & 79\\
  40 & 1.3579 &  $\pm$0.0129 &  .9197 &  $\pm$0.0032 & 0.0882 &
 $\pm$0.0005 & 0.0954 & $\pm$0.0005 & 0.0475 & $\pm$0.0005 & 0.0031 &
 $\pm$0.0005 & 57\\
  46 & 1.3871 &  $\pm$0.0045 &  .9227 &  $\pm$0.0043 & 0.0884 &
 $\pm$0.0007 & 0.0930 & $\pm$0.0007 & 0.0482 & $\pm$0.0007 & 0.0034 &
 $\pm$0.0008 & 56\\
  51 & 1.4061 &  $\pm$0.0061 &  .9329 &  $\pm$0.0083 & 0.0904 &
 $\pm$0.0007 & 0.0978 & $\pm$0.0007 & 0.0502 & $\pm$0.0007 & 0.0017 &
 $\pm$0.0006 & 76\\
  53 & 1.3507 &  $\pm$0.0210 &  .9270 &  $\pm$0.0097 & 0.0904 &
 $\pm$0.0008 & 0.0976 & $\pm$0.0008 & 0.0502 & $\pm$0.0009 & 0.0014 &
 $\pm$0.0006 & 76\\
  58 & 1.4131 &  $\pm$0.0055 &  .9374 &  $\pm$0.0083 & 0.0905 &
 $\pm$0.0007 & 0.0977 & $\pm$0.0007 & 0.0504 & $\pm$0.0007 & 0.0030 &
 $\pm$0.0008 & 77\\
  63 & 1.3799 &  $\pm$0.0095 &  .9456 &  $\pm$0.0063 & 0.0917 &
 $\pm$0.0006 & 0.0974 & $\pm$0.0006 & 0.0489 & $\pm$0.0006 & 0.0028 &
 $\pm$0.0007 & 66\\
  68 & 1.4028 &  $\pm$0.0053 &  .9011 &  $\pm$0.0040 & 0.0891 &
 $\pm$0.0007 & 0.0960 & $\pm$0.0007 & 0.0478 & $\pm$0.0007 & 0.0028 &
 $\pm$0.0007 & 74\\
  70 & 1.3892 &  $\pm$0.0066 &  .9109 &  $\pm$0.0046 & 0.0884 &
 $\pm$0.0009 & 0.0981 & $\pm$0.0010 & 0.0509 & $\pm$0.0010 & 0.0047 &
 $\pm$0.0012 & 63\\
  74 & 1.4092 &  $\pm$0.0034 &  .9193 &  $\pm$0.0029 & 0.0909 &
 $\pm$0.0006 & 0.0966 & $\pm$0.0006 & 0.0498 & $\pm$0.0006 & 0.0015 &
 $\pm$0.0006 & 76\\
  79 & 1.4194 &  $\pm$0.0064 &  .9453 &  $\pm$0.0040 & 0.0922 &
 $\pm$0.0006 & 0.0988 & $\pm$0.0006 & 0.0501 & $\pm$0.0007 & 0.0028 &
 $\pm$0.0008 & 72\\
  85 & 1.3948 &  $\pm$0.0070 &  .9136 &  $\pm$0.0050 & 0.0901 &
 $\pm$0.0007 & 0.0976 & $\pm$0.0008 & 0.0480 & $\pm$0.0008 & 0.0026 &
 $\pm$0.0009 & 74\\
  89 & 1.4072 &  $\pm$0.0091 &  .9086 &  $\pm$0.0088 & 0.0902 &
 $\pm$0.0007 & 0.0968 & $\pm$0.0007 & 0.0491 & $\pm$0.0007 & 0.0029 &
 $\pm$0.0008 & 76\\
  95 & 1.3673 &  $\pm$0.0035 &  .9179 &  $\pm$0.0038 & 0.0904 &
 $\pm$0.0006 & 0.0983 & $\pm$0.0006 & 0.0497 & $\pm$0.0006 & 0.0028 &
 $\pm$0.0006 & 44 \\
 100 & 1.3676 &  $\pm$0.0082 &  .9158 &  $\pm$0.0033 & 0.0928 &
 $\pm$0.0006 & 0.1000 & $\pm$0.0006 & 0.0497 & $\pm$0.0006 & 0.0023 &
 $\pm$0.0006 & 44 \\
 104 & 1.3731 &  $\pm$0.0037 &  .9205 &  $\pm$0.0028 & 0.0895 &
 $\pm$0.0007 & 0.1007 & $\pm$0.0007 & 0.0505 & $\pm$0.0007 & 0.0022 &
 $\pm$0.0007 & 44\\
 108 & 1.3480 &  $\pm$0.0057 &  .9174 &  $\pm$0.0032 & 0.0913 &
 $\pm$0.0007 & 0.0997 & $\pm$0.0007 & 0.0500 & $\pm$0.0007 & 0.0033 &
 $\pm$0.0010 & 44 \\ 
 112 & 1.4018 &  $\pm$0.0064 &  .9109 &  $\pm$0.0032 & 0.0896 &
 $\pm$0.0008 & 0.0974 & $\pm$0.0008 & 0.0482 & $\pm$0.0008 & 0.0020 &
 $\pm$0.0007 & 74 \\
 115 & 1.3663 &  $\pm$0.0052 &  .9220 &  $\pm$0.0031 & 0.0908 &
 $\pm$0.0007 & 0.0992 & $\pm$0.0007 & 0.0505 & $\pm$0.0007 & 0.0033 &
 $\pm$0.0008 & 44\\
 119 & 1.3938 &  $\pm$0.0053 &  .9147 &  $\pm$0.0043 & 0.0912 &
 $\pm$0.0007 & 0.0973 & $\pm$0.0007 & 0.0494 & $\pm$0.0007 & 0.0025 &
 $\pm$0.0006 & 78\\
 122 & 1.3714 &  $\pm$0.0114 &  .9280 &  $\pm$0.0088 & 0.0928 &
 $\pm$0.0007 & 0.0994 & $\pm$0.0007 & 0.0506 & $\pm$0.0007 & 0.0025 &
 $\pm$0.0007 & 44\\
 130 & 1.3726 &  $\pm$0.0035 &  .9236 &  $\pm$0.0025 & 0.0915 &
 $\pm$0.0006 & 0.1039 & $\pm$0.0007 & 0.0500 & $\pm$0.0006 & 0.0031 &
 $\pm$0.0007 & 38\\
 133 & 1.3966 &  $\pm$0.0032 &  .9144 &  $\pm$0.0027 & 0.0905 &
 $\pm$0.0004 & 0.0976 & $\pm$0.0004 & 0.0494 & $\pm$0.0005 & 0.0031 &
 $\pm$0.0006 & 69\\
 140 & 1.3464 &  $\pm$0.0131 &  .9212 &  $\pm$0.0039 & 0.0909 &
 $\pm$0.0006 & 0.0975 & $\pm$0.0006 & 0.0496 & $\pm$0.0006 & 0.0021 &
 $\pm$0.0006 & 63\\
 147 & 1.3991 &  $\pm$0.0116 &  .9131 &  $\pm$0.0043 & 0.0900 &
 $\pm$0.0007 & 0.0982 & $\pm$0.0007 & 0.0493 & $\pm$0.0007 & 0.0026 &
 $\pm$0.0007 & 57\\
 154 & 1.4964 &  $\pm$0.0472 &  .9133 &  $\pm$0.0076 & 0.0997 &
 $\pm$0.0008 & 0.1067 & $\pm$0.0008 & 0.0531 & $\pm$0.0008 & 0.0033 &
 $\pm$0.0009 & 72\\
 159 & 1.3155 &  $\pm$0.0098 &  .9257 &  $\pm$0.0036 & 0.0849 &
 $\pm$0.0007 & 0.0918 & $\pm$0.0007 & 0.0463 & $\pm$0.0008 & 0.0023 &
 $\pm$0.0007 & 79\\
 168 & 1.5046 &  $\pm$0.0224 &  .9933 &  $\pm$0.0216 & 0.0941 &
 $\pm$0.0009 & 0.1005 & $\pm$0.0009 & 0.0512 & $\pm$0.0006 & 0.0028 &
 $\pm$0.0006 & 79\\
 173 & 1.4648 &  $\pm$0.0209 &  .8518 &  $\pm$0.0087 & 0.0927 &
 $\pm$0.0008 & 0.0982 & $\pm$0.0008 & 0.0460 & $\pm$0.0017 & 0.0025 &
 $\pm$0.0009 & 74 \\
 175 & 1.4148 &  $\pm$0.0051 &  .9458 &  $\pm$0.0068 & 0.0917 &
 $\pm$0.0007 & 0.0984 & $\pm$0.0007 & 0.0509 & $\pm$0.0010 & 0.0024 &
 $\pm$0.0006 & 78\\
 186 & 1.3729 &  $\pm$0.0083 &  .8128 &  $\pm$0.0274 & 0.0904 &
 $\pm$0.0006 & 0.0984 & $\pm$0.0007 & 0.0493 & $\pm$0.0008 & 0.0019 &
 $\pm$0.0007 & 44\\
 189 & 1.3928 &  $\pm$0.0047 &  .9776 &  $\pm$0.0102 & 0.0895 &
 $\pm$0.0006 & 0.0953 & $\pm$0.0006 & 0.0498 & $\pm$0.0007 & 0.0030 &
 $\pm$0.0006 & 68\\
 192 & 1.3564 &  $\pm$0.0057 &  .9497 &  $\pm$0.0104 & 0.0903 &
 $\pm$0.0007 & 0.0985 & $\pm$0.0007 & 0.0489 & $\pm$0.0006 & 0.0033 &
 $\pm$0.0007 & 44\\
 197 & 1.3875 &  $\pm$0.0094 &  .9269 &  $\pm$0.0031 & 0.0863 &
 $\pm$0.0007 & 0.0925 & $\pm$0.0007 & 0.0488 & $\pm$0.0006 & 0.0000 &
 $\pm$0.0000 & 29\\

\end{tabular}
\end{center}
\caption{ Flux densities (in Jy) of the observed sources and their
   errors at 15~GHz. The first column refers to the days starting
  from 1994 March 3. The last column gives the elevation of 
  B1422+231 in degrees. The flux densities of the two calibrators, 3C287
  and OQ208 were determined from the program GETJY while that of
  lensed images were determined by fitting a gaussian to the image.
  }
\end{table*}

The data for B1422+231 was imaged using standard procedure in 
{\sc AIPS}. Typical rms noise levels in the maps were about 0.3~mJy/beam
at 8.4~GHz and 0.4~mJy/beam at 15~GHz.  Fig. 1 shows a 15~GHz map of
B1422+231 where the images have been labelled. The flux densities were
measured from the maps by fitting gaussians to each of the four images
using the program JMFIT.  We quote the errors in flux densities as
given by this program. The flux densities of the calibrators and of
the four lensed images are given in Table 1 and these are plotted in
Fig.2 (for 8.4~GHz results) and Table 2 and Fig. 3 (for 15~GHz results).

\begin{figure}
\raisebox{-3.2cm}
{\begin{minipage}{0.3cm}
\mbox{}
\parbox{0.3cm}{}
\end{minipage}}
\begin{minipage}{7cm}
\mbox{}
\psfig{figure=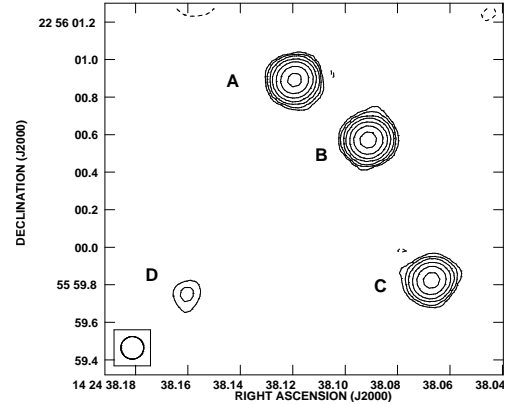,width=6.5cm,rheight=6.5cm,rwidth=6.5cm}
\centering
\end{minipage}
\caption{15~GHz map of B1422+231. The map has been restored with a
  circular gaussian with FWHM of 0.2~arcsec, drawn at the bottom left
  hand corner. The contour levels are
  $-2,-1,1,2,4,8,16,32,64$~mJy beam$^{-1}$. Negative contours are
  dashed. The rms noise in the map is 0.37~mJy beam$^{-1}$. The images
  have been labelled.
  }
\end{figure}

\begin{figure*}
\raisebox{-3.2cm}
{\begin{minipage}{0.3cm}
\mbox{}
\parbox{0.3cm}{}
\end{minipage}}
\begin{minipage}{12cm}
\mbox{}
\psfig{figure=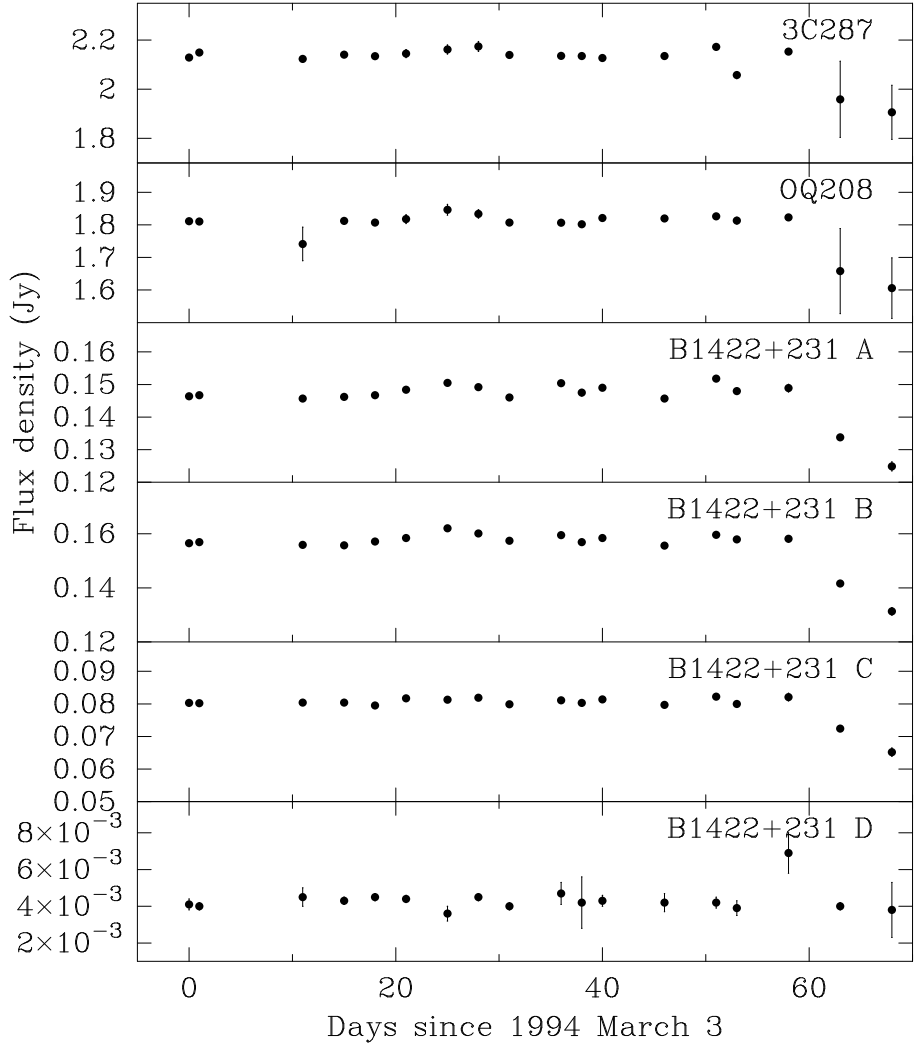,width=12cm,rheight=12cm,rwidth=12cm}
\centering
\end{minipage}
\vspace{2.5cm}
\caption{ Light curves for the observed sources at 8.4~GHz. Source name is
  indicated in the respective panel. 
           }
\end{figure*}

\begin{figure*}
\raisebox{-3.2cm}
{\begin{minipage}{0.3cm}
\mbox{}
\parbox{0.3cm}{}
\end{minipage}}
\begin{minipage}{12cm}
\mbox{}
\psfig{figure=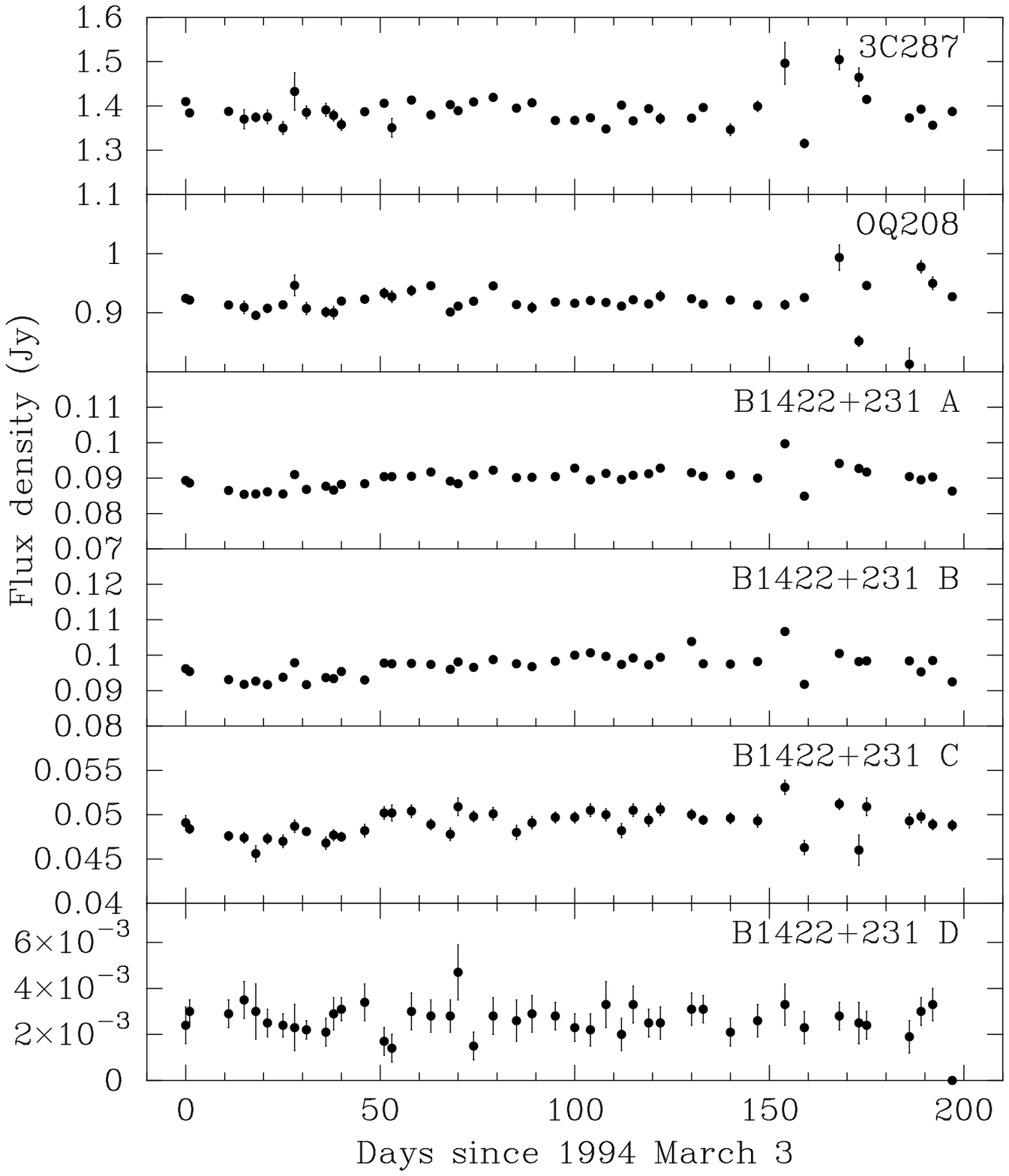,width=12cm,rheight=12cm,rwidth=12cm}
\centering
\end{minipage}
\vspace{2.5cm}
\caption{Light curves of the observed sources at 15~GHz. Source name
  is indicated at the respective panel.   
}
\end{figure*}

In what follows we chose not to analyse the results of 8.4~GHz due to
insufficient points in the light curve. We concentrate on the 15~GHz
results.

It is evident that any correlated variability (at the same epoch) in
the light curves of calibrators as well as the target sources is due
to errors in calibration.  We, therefore, normalised each of the light
curves of the sources by dividing it by the flux density of its first
epoch. Then we divided these normalised points by the corresponding
point of OQ208 from the light curves of all the sources in order to
remove the calibration errors. The rms variation in the light curve of
3C287 improved from 2.5\% to 1.7\% after this division (Fig. 6).  

Figure 4 shows the light curves of A, B and C images of B1422+231.
Figure 5 shows the light curve of image D.  It is clear from Fig. 4
that we do detect variability of the three brighter images of
B1422+231. Image D is rather weak and suffers from larger
uncertainties due to poor signal-to-noise ratio and hence has
correspondingly larger error bars.  It is, therefore, not possible to
conclude about its variability from the present observations.  For the
others, we observe peak-to-peak variability of about 5\% during our
observing run of 197 days.

\section{New method for Determination of Time--delay}

In the past, the conventional method to cross-correlating the fluxes
of the images to determine the time delay between them has not been
very successful when the noise at the time period of expected time
delay was non-negligible, even though the overall signal to noise
ratio of the data could be very good. This was apparent in the case of
$0957+561$ where time delay was around 417 days while some of the
microlens events had similar duration (Schild \& Thomson 1997,
Kundi\'c et al.  1997b, Haarsma et al. 1999).  This problem is severe
for B1422+231 due to the low amplitude of variability in radio.

Our method is based on identifying a single component of the intrinsic
variability and using this component to cross correlate between the
images (cf. Narasimha 2000). In the conventional methods, attempts are
usually made to cross--correlate all the frequencies of variability
simultaneously, and hence the method is most sensitive to the highest
frequency. Unfortunately, the variability in the data at the highest
frequency is the noise. We attempt to derive an alternative approach,
in which, essentially one single frequency is analysed and the noise
in the phase of this component is minimised before the feature is used
for cross--correlation between the images.  A smoothed cubic spline is
used for the identification of the feature. We do not use any specific
form for the variability, but in view of its application for many
similar systems we shall illustrate and explain the general method
through an ideal example where the term phase difference is easily
understood.

Let the smooth component of flux obtained from the observed sample be
represented by a function of the form,

$$f_\nu(t) = f_o + A sin \left[\omega \left(t-t'\right) + \phi\right] $$

\noindent
Here $f_o$ is the steady flux, $A$ is the amplitude of variability,
$\omega$ is the frequency of intrinsic variability of the signal of
interest and the crucial term required for computation of the time
delay is the observed phase factor $\phi$ for the different images.

If the external noises were absent and the sampling were sufficiently
good, the flux would be

$$F_\nu(t) = F_o + \alpha sin \left[\omega_o \left(t-t'\right) + \phi_o \right] $$

\noindent
We assume that $\alpha$, $\omega_o$ and $\phi_o$ are close to the
observed sample values $A$, $\omega$ and $\phi$ respectively.  Then, a
Taylor expansion for the sample values about the population mean is
valid. Carrying out a $\chi^2$ minimisation we can obtain four linear
relations for the four unknowns. If the observations span one cycle of
variability or more, the mean values for terms like
sin($\omega t + \phi$), cos($\omega t + \phi$)
summed over the observed epochs will
tend to zero. Consequently, we can select a value of $t'$ close to the
mean epoch of the observation such that the computed sample values of
$A$, $\omega$, $f_\nu$ and $\phi$ will be stochastically independent.
The variance of $\phi$ can, then, be approximated by

$$Variance(\phi) = 2/ \sum_{i} { (f_{\nu,i} - \bar{f}_\nu)^2 \over {\sigma_i^2} } $$

The time-delay between the images can be estimated from the
cross-correlation between the variable part of the fluxes obtained from
the smoothed cubic splines. The cross correlation will have the form

$$ R = cos\left(\omega \tau + \phi_1 - \phi_2 \right) ,$$

\noindent
where $\tau$ is the assumed time delay between the images.  The phase
factors $\phi_1$ and $\phi_2$ refer to the variable components of the
smoothed cubic spline fits to the observed fluxes in the images 1 and
2.  The errors in the data points are contributed almost entirely by
the calibration source and the steady part of the signal. The time
period over which the source is monitored is very long compared to the
time--delay. Consequently, the variance in R can be approximated, to a
good extent, by the product of the variance of the normalised flux of
the two images.  As we described earlier, since we are interested only
in the phase factor of the variable component of the fluxes, the
errors in the normalised smoothed flux represent the errors in the
computed phase factor.  The variance is an indicator of the expected
error in the computed flux and hence, the estimated time--delays.  If
the smoothed flux has exactly the same form in the images, the
correlation will have a peak value of unity, but in practice it will
be less when the fitting is done independently.  In view of the form
of the expression for R, the error in the computed time delay is the
time interval over which the correlation, R, drops by half the
estimated variance of the phase $\phi$ of the smoothed fit to the
flux.

\section{Results }

For the system B1422+231, the 15~GHz observations were spread over 197
days and a variability of 5.2\% was detected. We estimate the period
of variability to be about 216 days.  The observations were carried
out at intervals of 2 to 11 days. It is reasonable to assume that the
observations lasted for one cycle of variation. For the 8.4~GHz, the
observations lasted for less than seventy days, number of epochs were
less.  Consequently, we are not able to carry out the analysis
discussed here.  Hence our time--delay analysis is restricted to the
15~GHz data.

\begin{figure}
\raisebox{-3.2cm}
{\begin{minipage}{0.3cm}
\mbox{}
\parbox{0.3cm}{}
\end{minipage}}
\vspace{1.0cm}
\begin{minipage}{7cm}
\mbox{}
\psfig{figure=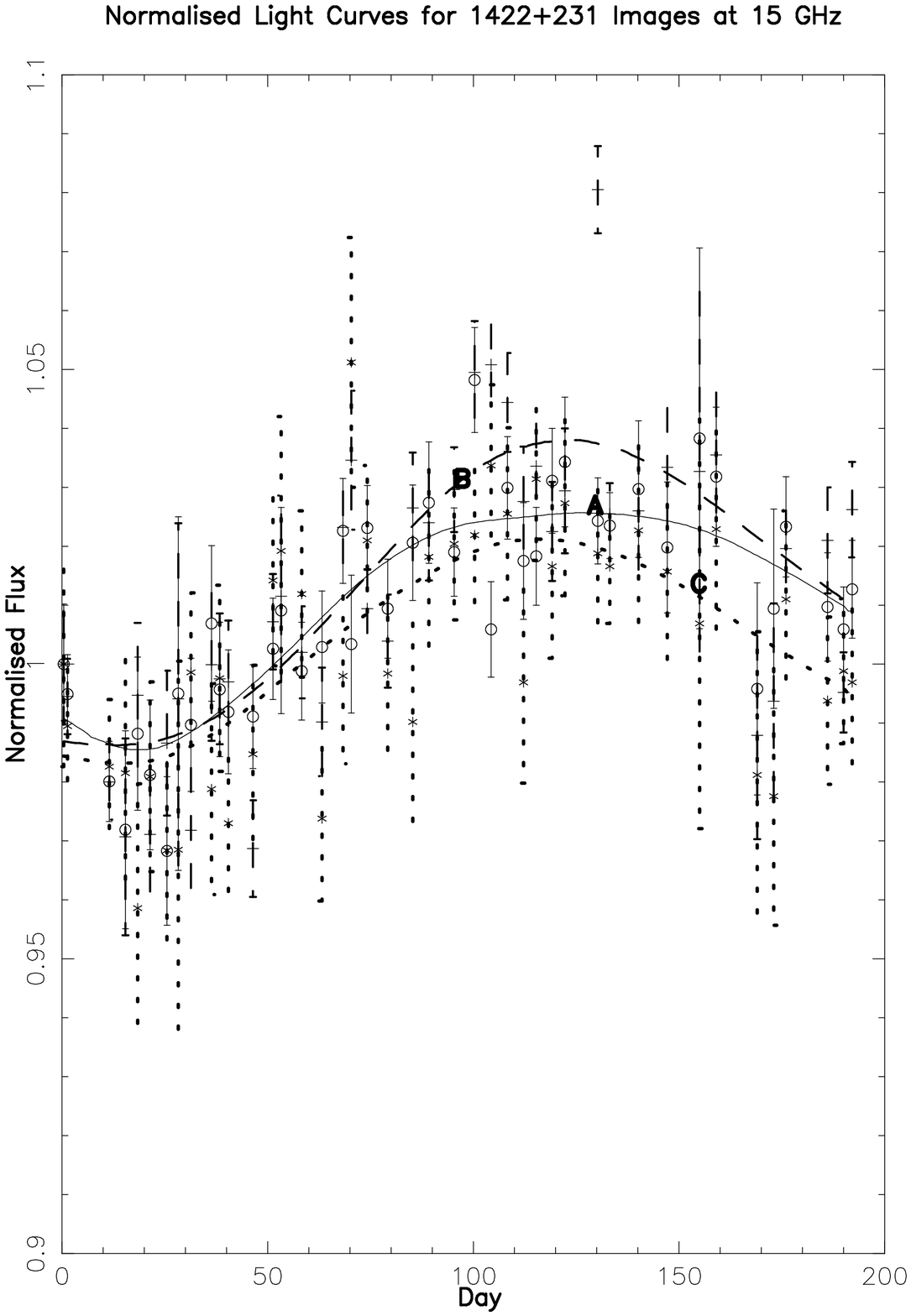,width=6.5cm,rheight=6.5cm,rwidth=6.5cm}
\centering
\end{minipage}
\vspace{2.0cm}
\caption{
  Smoothed light curves (spline fit) of images A ($\circ$), B ($+$)
  and C ($\star$) of B1422+231 at 15~GHz. The light curves have been
  normalised. Observations for individual epochs are marked
  with 1$\sigma$ error bars. The X-axis refers to the days since
  1994 March 3.
  }
\end{figure}

\begin{figure}
\raisebox{-3.2cm}
{\begin{minipage}{0.3cm}
\mbox{}
\parbox{0.3cm}{}
\end{minipage}}
\vspace{1.0cm}
\begin{minipage}{7cm}
\mbox{}
\psfig{figure=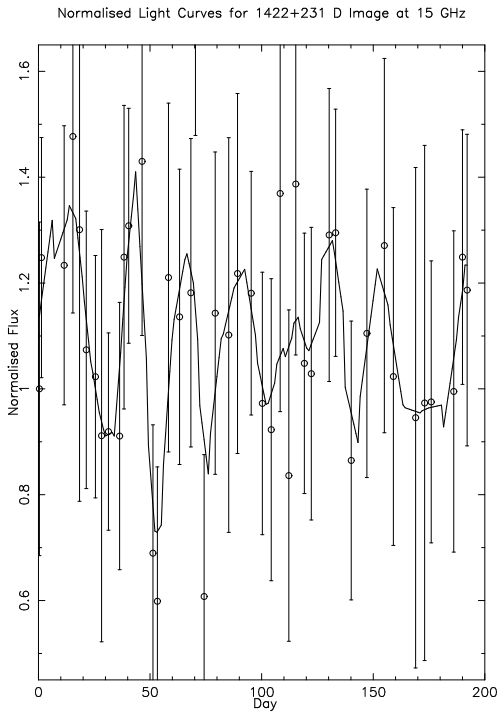,width=6.5cm,rheight=6.5cm,rwidth=6.5cm}
\centering
\end{minipage}
\vspace{2.0cm}
\caption{Normalised light curve of image D of B1422+231 at 15~GHz. The
  X-axis refers to the days since 1994 March 3. The solid line is the
  spline fit.
  }
\end{figure}

\begin{figure}
\raisebox{-3.2cm}
{\begin{minipage}{0.3cm}
\mbox{}
\parbox{0.3cm}{}
\end{minipage}}
\vspace{1.0cm}
\begin{minipage}{7cm}
\mbox{}
\rotate[r]{\psfig{figure=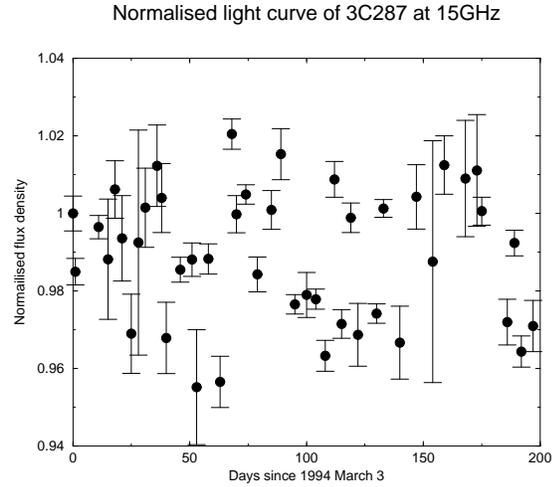,width=6.5cm,rheight=6.5cm,rwidth=6.5cm}}
\centering
\end{minipage}
\caption{ Normalised light curve of the calibrator 3C287 after
  dividing it by that of OQ208. No systematic variability is observed
  in this light curve.
}
\end{figure}

\begin{figure}
\raisebox{-3.2cm}
{\begin{minipage}{0.3cm}
\mbox{}
\parbox{0.3cm}{}
\end{minipage}}
\vspace{1.0cm}
\begin{minipage}{7cm}
\mbox{}
\psfig{figure=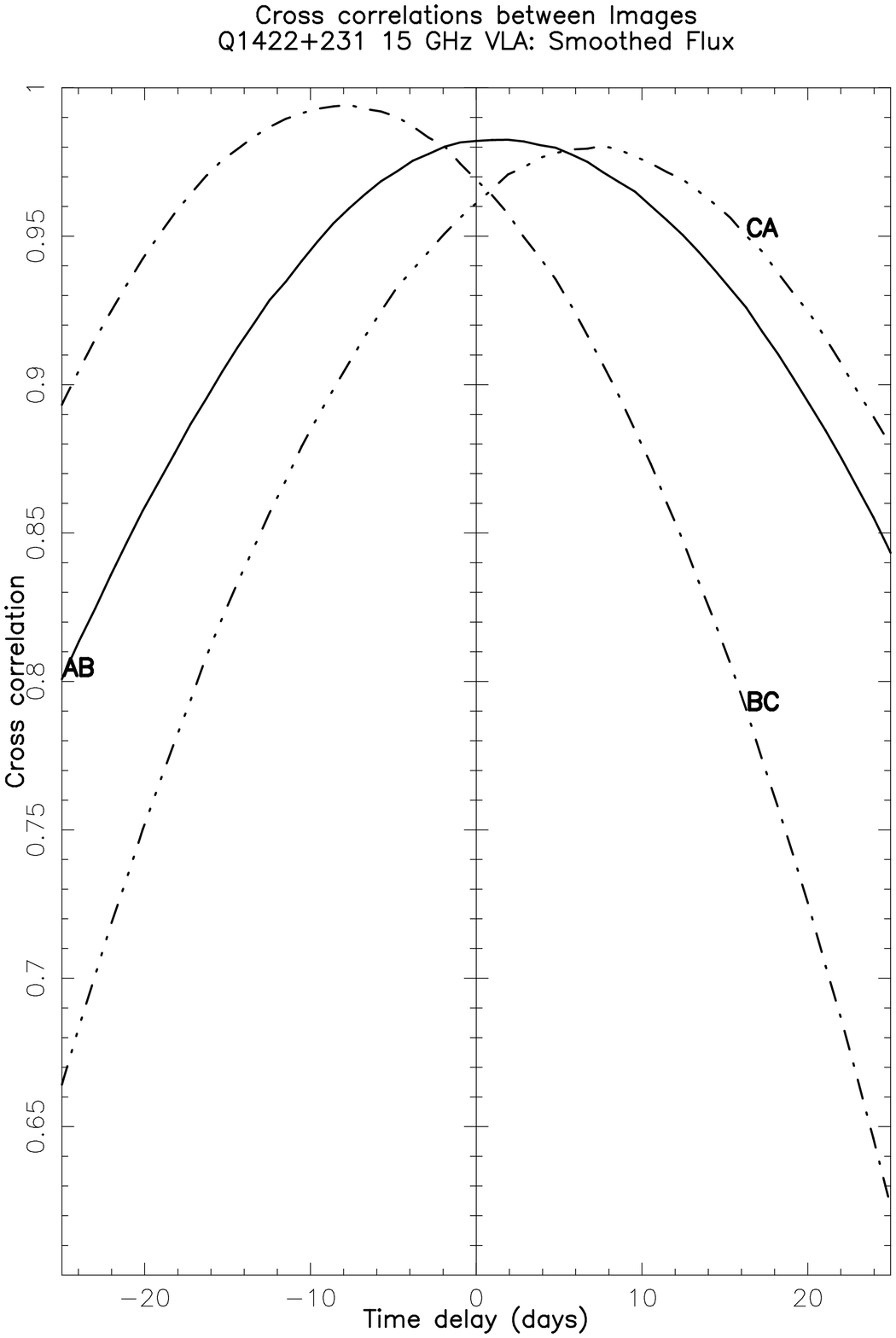,width=6.5cm,rheight=6.5cm,rwidth=6.5cm}
\centering
\end{minipage}
\vspace{2.5cm}
\caption{Cross correlation between the images of B1422+231 at 15~GHz
  using the smoothed light curve as explained in the text.
}
\end{figure}

The variances of $\phi$ for the images B, A and C are at 0.013 level
which is equivalent to 4 days (variance of 0.01 is equivalent to an
error of 0.1~radian, and as the period is about 216 days, this amounts
to about 4 days).  This is a consequence of the time interval between
the observations being at 2 to 11 days.  Since the time intervals
between observations are not fixed, we are able to separate a
time--delay of less than the minimum time interval between the
successive observations.  But our computed error is rather high which
shows that with a better sampling of the observations we should be
able to get time--delays for all the images.  In practice, the
uncertainty in the relative fluxes of the images is mainly due to the
day to day calibration error rather than the
relative flux of the images with respect to the calibrator. Hence,
strictly, the stochastic independence of the data cannot be claimed.
Nevertheless, we treat the normalised fluxes of images to be
stochastically independent.  This results in higher error estimate.

However, for the image D, the error in individual observations is
considerable due to the faintness of the image.  Consequently, the
estimated variance of the phase is around 7, corresponding to an
uncertainty of over 150 degrees, implying that the relative phase of
image D with respect to other images cannot be estimated from the
given sample. So, we have not carried out any analysis for this image,
in spite of its significance for cosmological purposes.  The derived
time delays between images B and A is 1.5$\pm$1.4d, between A and C is
7.6$\pm$2.5d and between B and C is 8.2$\pm$2.0d.

When is this method superior to direct estimation of the time delay?
If the smoothed cubic spline can isolate a single component of the
intrinsic flux variability from the mean flux of the
images, and any contribution due to micro-lensing is filtered from the
smooth function, then the signal in the correlation will improve
compared to the noise by a factor of (n-4), where n is the number of
epochs of observation and the number 4 is based on the minimum number
of data points required for the computations of the smoothed spline.
In practice, the errors between the various images are not independent
because as in the case of our data for B1422+231, the main
contribution to error in the smoothed fit is from the calibrating
sources which are common to all the images.  Consequently, our error
estimate is probably conservative; the time delays could be closer to
our estimations than the error bars suggest. This can be seen also
from the fact that the sum of the time delays between images
BA and between AC agrees well with that of BC.

\section{Discussion}

The time delays between the images A, B and C have been determined,
though not very accurately. However, the monitoring of the weaker
image D has not been successful mainly because the fractional errors
in the flux densities were considerably larger than the amplitude of
the intrinsic variability.  For the distance estimate, the time delay
between image D and other images is important because it is a measure
of the lens mass.  Hence our results, so far, do not provide any
reliable distance determination. It would be necessary to carry out
monitoring for a longer time with considerably better precision in
order to estimate the time delay between image D and the brighter
images.

The models of Narasimha \& Patnaik (1993) predicted a value of time
delay between images A and C of the order of 9 days for a distance to
the lens of 1000 Mpc (signal appearing in C first), assuming a single
component lens and H$_0$ to be 50~km~s~$^{-1}$Mpc$^{-1}$. Our
measurement of AC time delay of 7.6$\pm$2.5d implies a Hubble constant
in the range of 64 to 75~km~s$^{-1}$~Mpc$^{-1}$ for the standard
cosmology, with only a weak dependence on $\Omega$ and $\Lambda$.
Narasimha \& Srianand (1999) considered abstract models comparing the
caustic structure near the images A, B and C using the flux ratios in
the emission lines, optical and radio continuum. Their result cannot
be directly translated into time delay, but if the size of Lyman
emission region is less than $10^{17}$ cm, a large scale length of the
lens is implied and the Hubble constant turns out to be less than
35~km~s$^{-1}$~Mpc$^{-1}$.  Since there is evidence for a weak group
of galaxies near the main lens, the result is expected to be in
between the two limits discussed above.

However, a natural question to address is that whether the method
suggested here is reliable and has advantages over the conventional
data analysis techniques.  We shall briefly address these questions by
showing our results and some comments for the three systems, namely,
B1608+656, B0218+357 and PG1115+080, for which time delays have been
available during the last few years, but the data are still not good
compared to 0957+561.

\section{Variability and Time--delay in other lens systems}

We have analysed the available data for the systems B1608+656
(Fassnacht et~al., 1999), B0218+357 (Biggs et~al., 1999) and
PG1115+080 (Schechter et~al., 1997).  We have displayed one typical
diagram for the time delay based on polarization angles in the system
B0218+357 (Fig. 8) and a summary of the results for the three
systems is given in Table 3. 

The data for B1608+656 are very good, in spite of low amplitude of
the variability and multi-component fluctuations.
The excellent calibration based on stable reference source
and elimination of the questionable data make
the derived time delay reliable.
Our results for the three combination of the time delays
(AB, BC and BD) are in good agreement with the value derived
by the observers' team, and the value for AB is very close to
the differential time delay BC-AC. However, it would be
worth pointing out that the smoothed spline cross--correlation
peaks at much lower than 1, which indicates that there could be
considerable amount of noise introduced en route in all the images.
Possibly multi-frequency observations and polarization
data would be required to assess the nature of this noise.

\begin{figure}
\raisebox{-3.2cm}
{\begin{minipage}{0.3cm}
\mbox{}
\parbox{0.3cm}{}
\end{minipage}}
\vspace{1.0cm}
\begin{minipage}{7cm}
\mbox{}
\rotate[r]{\psfig{figure=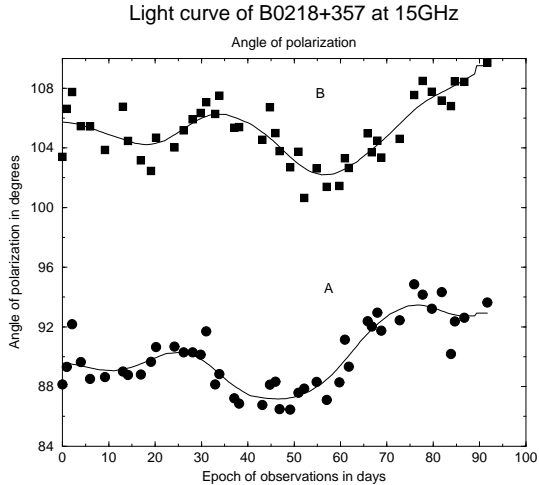,width=6.5cm,rheight=6.5cm,rwidth=6.5cm}}
\centering
\end{minipage}
\vspace{0cm}
\caption{
Smoothed curve (spline fit) through the observed data (position angle of
polarization) of B0218+357 at 15GHz. Light curves for images A and B
are labelled. Data
are taken from Biggs et~al. (1999).
  }
\end{figure}

\begin{figure}
\raisebox{-3.2cm}
{\begin{minipage}{0.3cm}
\mbox{}
\parbox{0.3cm}{}
\end{minipage}}
\vspace{1.0cm}
\begin{minipage}{7cm}
\mbox{}
\rotate[r]{\psfig{figure=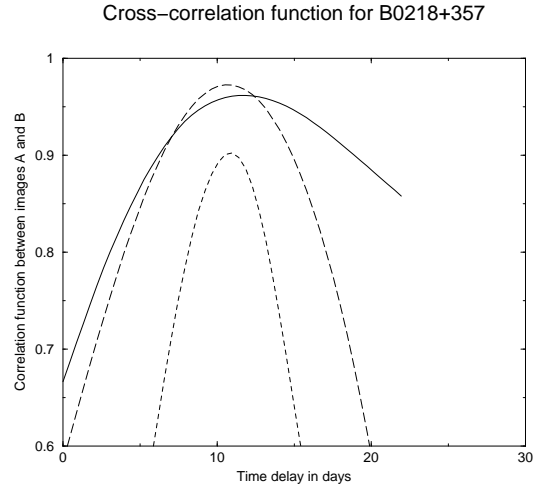,width=6.5cm,rheight=6.5cm,rwidth=6.5cm}}
\centering
\end{minipage}
\vspace{0cm}
\caption{
Cross-correlation function for the light curve of B0218+357. Solid
line represents the correlation function for the total intensity,
long dashed curve represents that for the position angle of
polarization, and small dashed curve represents that for the
fractional polarization of B0218+357 at 15GHz. Data are taken from
Biggs et~al. (1999).
  }
\end{figure}

For the system B0218+357, at the higher frequency, all the three
methods, namely, total flux, polarization fraction and the the
polarization angle give nearly identical results (Fig. 9 and Table 3),
all in agreement with the values given by the observers' team.  The
polarization angle appears to have the highest correlation between the
images in our method. We think that polarization
angle is among the most trustworthy indicators of the intrinsic
variability, as pointed out by Nair et~al. (1993) and Nair (1994).

\begin{table*}
\begin{center}
{\small
\caption {\it Comparison of time--delay analysis
for three lens systems:} The method developed here is
used for the recent data available
for three lenses and the results are compared.
}
\medskip
\begin{tabular}{lllll}\\
System & Images & Observers            & Present Technique  & Reference \\
       &        & Time-delay (days)    & Time-delay (days)  & \\
       &        &                      &                    & \\
\hline
B1608+656 & $\Delta t_{BA}$ & 31$\pm$7 & 30.4$\pm$4.5 & Fassnacht et~al. (1999) \\
          & $\Delta t_{BC}$ & 36$\pm$7 & 37.6$\pm$4.5        &      \\
          & $\Delta t_{BD}$ & 76$^{+9}_{-10}$ & 75.3$\pm$4.2 &      \\
          & $\Delta t_{AC}$ &                 & 4.5$\pm$4.5  &      \\
\hline
B0218+357 &                 &                 &               & Biggs et~al. (1999) \\
15~GHz:    &                 &                 &               &  \\
         Flux & $\Delta t_{AB}$ & 10.6$^{+0.7}_{-0.5}$ & 11.6$\pm$2 & \\
 Polarization fraction &  & 11.4$\pm$0.3        & 11.4$\pm$1      & \\
 Polarization angle &    & 10.2$^{+0.3}_{-0.4}$ & 11.2$\pm$2      & \\
8.4~GHz:       &          &                      &                 & \\
        Flux  &          & 10.1$^{+1.4}_{-0.7}$ & $^{\ddagger}$   & \\
 Polarization fraction &  &                      & 14.4$\pm$3      & \\
 Polarization angle &    &                      & 8.4$\pm$3       & \\
\hline
PG1115+080    & $\Delta t_{CB}$ & 23.7$\pm$3.4 & 19.6$\pm$5.4 & Schechter et~al.(1997) \\
              & $\Delta t_{CA}$ & 9.4$\pm$3.4  & 12.4$\pm$4.8 & \\
              & $\Delta t_{AB}$ & 14.3$\pm$3.4 & 6.7$\pm$4.8  & \\
              &                 &              &              & \\
$^{\ast}$     & $\Delta t_{CB}$ &              & 24.4$\pm$4.8 &  \\
              & $\Delta t_{CA}$ &              & 13.6$\pm$4.2 &  \\
              & $\Delta t_{AB}$ &              & 10.3$\pm$4.2 & \\
\hline
\end{tabular}
\end{center}
\medskip
\noindent
$^{\ast}$\ After eliminating two days when the reference star
exhibited considerable fluctuation. \\
$^{\ddagger}$\ The smoothed cubic spline did not converge
and hence we cannot get a time--delay for the total flux. \\
\end{table*}

For the system PG1115+080, again, our results are broadly in agreement
with Schechter et~al. (1997). However, a few comments are worth
mentioning.  First, the time delay between images B and C are
generally stable against analysis technique or some amount of noise in
the data. But the time--delay between the merging image A and the
other two are sensitive to the method adopted as well as the noise in
the reference star.  Second, for the whole data, our time--delay
between images B and C is slightly smaller than the value given by
Schechter et al. But if we drop two days on which the fluctuation in
the magnitude of the reference star was comparable to the amplitude of
variability of the quasar, the time--delay becomes 24.4 days, which is
exactly in between the values derived by Schechter et~al. and Barkana
(1997).  Third, our value for the time delay between images A and B
range from 6 to 11 days, depending on using or dropping the data for
the two days.  The corresponding value for the time--delay between
images C and A vary between 14 and 13 days.  Consequently, our
time--delay ratio between CA and AB are always consistent with the
lens models; the small ratio supports a compact lens and higher value
for the Hubble constant while the larger ratio favours an extended
lens and smaller Hubble constant.  

\section{Conclusions}

We have monitored B1422+231 with the VLA at two different frequencies.
The light curves were sampled at intervals of 2 to 10 days only
because we did not have a good idea of the amplitude of intrinsic
variability and were interested in the image D in view of its
importance for cosmological distance calibration.  Consequently, we
are unable to get very accurate value for the time delay between
images.  But we are able to determine the time delay between the
bright images using the 15~GHz observations, obtaining 1.5, 7.6 and
8.2 days for AB, CA and CB time delays respectively.  More extensive
monitoring at significantly higher signal to noise ratio is necessary
to determine the time delay between the weak image D and the other
images.

\subsection*{ACKNOWLEDGMENTS}

We would like to thank J.~Schmid-Burgk and K.M.~Menten for critical
comments and the comments of an anonymous referee which resulted in
improving the presentation of the paper.  We have made use of
information available at the CASTLE web-site
(http://cfa-www.harvard.edu/glensdata/) for gravitational lenses
maintained by C.S. Kochanek, E.E. Falco, C. Impey, J. Leh{\'a}r, B. McLeod
and H.-W. Rix. We are grateful to NRAO for their support in carrying
out these observations.  The National Radio Astronomy Observatory is a
facility of the National Science Foundation operated under cooperative
agreement by Associated Universities, Inc.  DN acknowledges support
from Indo-French Centre for the promotion of Advanced Research,
through contract 1410-2.


\def\beginrefer{\section*{References}%
\begin{quotation}\mbox{}\par}
\def\refer#1\par{{\setlength{\parindent}{-\leftmargin}\indent#1\par}}
\def\endrefer{\end{quotation}}

\beginrefer

\refer Akujor, C.E., Patnaik, A.R., Smoker, J.V., Garrington, S.T.,
1996, in Kochanek, C.S, Hewitt, J.N., eds,  Proceedings of the 173rd
Symposium of the IAU `Astrophysical applications of gravitational
lensing', Kluwer Academic Publishers, Dordrecht, p335 \\
\refer Barkana, R., 1997, ApJ, 489, 21 \\
\refer Bechtold, J., Yee, H.K.C.,  1995, AJ, 110, 1984\\
\refer Biggs, A.D., Browne, I.W.A., Helbig, P., Koopmans, L.V.E.,
Wilkinson, P.N., Perley, R.A., 1999, MNRAS, 304, 349 \\
\refer Fassnacht, C.D., Pearson, T.J., Readhead, A.C.S., Browne,
I.W.A., Koopmans, L.V.E., Myers, S.T., Wilkinson, P.N., 1999,
ApJ, 527, 498 \\
\refer Haarsma, D.B., Hewitt, J.N., Leh{\'a}r, J., Burke, B.F., 1999, ApJ, 510,
64 \\
\refer Hogg, D.W., Blandford, R.D., 1994, MNRAS, 268, 889 \\
\refer Impey, C.D., Foltz, C.B., Petry, C.E., Browne, I.W.A., Patnaik, 
A.R., 1996, ApJ, 462, L53\\
\refer Koopmans, L.V.E., de Bruyn, A.G., Xanthopolous, E., Fassnacht,
C.D., 2000, AA, in press \\
\refer Kormann, R., Schneider, P., Bartlemann, M., 1994, A\&A, 286,
357 \\
\refer Kundi\'c, T., Hogg, D.W., Blandford, R.D., Cohen, J.G., Lubin,
L.M., Larkin, J.E., 1997a, AJ, 114, 2276\\
\refer Kundi\'c, T., Turner, E.L., Colley, W.N., Gott, III,R., Rhoads
, J.E., Wang, Y., Bergeron, L.E., Gloria, K.A., Long, D.C., Malhotra,
S., Wambsganss, J.,  1997b,  ApJ, 482, 75\\
\refer Lawrence, C.R., Neugebauer, G., Weir, N., Matthews, K., Patnaik,
 A.R., 1992, MNRAS, 259, 5P\\
\refer Lovell, J., Jauncey, D.L., Reynolds, J.E., Wieringa, M.H.,
King, E.A., Tzioumis, A.K., McCulloch, P.M., Edwards, P.G.,
1998, ApJ, 508, L51 \\
\refer Mao, S., Schneider, P., 1998, MNRAS, 295, 587\\
\refer Nair, S., 1994, Ph.D. Thesis, Bombay University, unpublished \\
\refer Nair, S. Narasimha, D., Rao, A.P., 1993, ApJ, 407, 46 \\
\refer Narasimha, D., 1999, Pramana, 53, 921 \\
\refer Narasimha, D., 2000, in Rencontres de Moriond, Les Arcs, eds.  Y. Mellier, J.-P. Kneib \& M. Moniez \\
\refer Narasimha, D., Patnaik, A.R., 1994, in Surdej, J.,
Fraipont-Caro, D, Grosset, E., Refsdal, S., Remy, M., eds, Proc. 31st
Li{\`e}ge International Astrophysical Colloq.,  Gravitational
Lenses in the Universe, Universit{\'e} de Li{\`e}ge, Belgique, p.295\\
\refer Narasimha, D., Srianand, R., 1999, in Bulletin of Astronomical
Society of India (in press) \\
\refer Patnaik, A.R., Browne, I.W.A., Walsh, D., Chaffee, F.H., Foltz, 
C.B.,  1992, MNRAS, 259, 1P \\
\refer Patnaik, A.R., Kemball, A.J., Porcas, R.W., Garrett, M.A.,
1999, MNRAS, 307, L1 \\
\refer Refsdal, S., 1964, MNRAS, 128, 307 \\
\refer Remy, M., Surdej, J., Smette, A., Claeskens, J.-F.,  1993, AA,
278, L19 \\
\refer Schild, R., Thomson, D.J.,  1997, AJ, 113, 130 \\
\refer Schechter, P.L., Bailyn, C. D., Barr, R., Barvainis, R., Becker, C. M.,
  Bernstein, G. M., Blakeslee, J. P., Bus, Schelte J., Dressler, A., Falco, E. E.,
  Fesen, R. A., Fischer, P., Gebhardt, K., Harmer, D., Hewitt, J. N.,
  Hjorth, J., Hurt, T., Jaunsen, A. O., Mateo, M., Mehlert, D.,
  Richstone, D. O., Sparke, L. S., Thorstensen, J. R., Tonry, J. L.,
  Wegner, G., Willmarth, D. W., Worthey, G., 1997, ApJ, 475, 85 \\
\refer Tonry, J.L., 1998, AJ, 115, 1\\
\refer Yee, H.K.C., Bechtold. J., 1996, AJ, 111, 1007 \\
\refer Yee, H.K.C., Ellingson, E.,  1994, AJ, 107, 28 \\

\endrefer

\bsp

\end{document}